# Designing Mobile Interaction Guidelines to Account for Situationally Induced Impairments and Disabilities (SIID) and Severely Constraining Situational Impairments (SCSI)

**Sidas Saulynas**
UMBC and Stevenson University
Baltimore, MD, USA
saulyn1@umbc.edu

**Ravi Kuber**
UMBC
Baltimore, MD, USA
rkuber@umbc.edu

**ABSTRACT**

This research investigates the variety and complexity of situational impairment events (SIE) that are being experienced by users of smartphone technology of all abilities. The authors have created a classification system to help describe the different types of SIE as well as differentiate a certain sub-group of events that were identified as severely constraining. Continuing research examined workarounds that users deploy when attempting to complete a mobile I/O transaction in the presence of an SIE, as well as social/cultural barriers to attempting mobile interaction that users recognize but do not always follow. The ultimate goal of this research arc would be the creation of guidelines to assist mobile designers and researchers in the accounting of SIE and perhaps different design considerations for those events deemed severely constraining.

**KEYWORDS**

Situational Impairments; Situationally Induced Impairments and Disabilities; SIID; Severely Constraining Situational Impairments; SCSI; Mobile Interaction; Guidelines.

**INTRODUCTION AND RELATED WORK**

The ability to access information and conduct I/O transactions on the go has added tremendous value to our increasingly complex lives. For example, if one was with a group in a crowded location and became separated, prior to the omnipresence of mobile interaction, this could lead to delays, frustration, and even anxiety particularly if one of the separated parties was a small child. The existence of mobile technology in this situation, affords the ability not only to SMS "Where are you?" but to incorporate multimedia to enhance the information richness of the communication channel (e.g. An SMS response, "I am under this sign" + a digital image of the sign.)

The same tool, however, that brings a sense of order and control to a chaotic situation, could equally produce greater chaos and disorder. For example, the attempted communication with the group is disrupted by other simultaneous exogenous communications (e.g. SMS interrupted by a phone call). Also, the inability to access the device to send or receive communication due to unavailable input resources (i.e. hands full), limited device resources (e.g. low battery or weak signal), or other aspects of the environmental context (e.g. too bright to see screen, too loud to send/received voice, or concern of possible device theft in a crowded and unfamiliar environment) can all conceivably lead to, at least temporarily, a net increase in chaos.

The rapid adoption and omnipresent use of mobile devices seems to have produced a new complexity by-product in the "Situationally Induced Impairments and Disabilities (SIID)" [8] problem space. This by-product has been termed "Severely Constraining Situational Impairments (SCSI)" or "an occurrence of a situational impairment and disability where a workaround is not available or easily obtained, or where a technological solution was found that only led to the introduction of a new situational impairment and disability" [6]. SIIDs, and SCSIs in particular, represent a real accessibility challenge for mobile device use *on the go*, and the effects of these challenges can often be dangerous.

**Related Work**

Mobile interaction represents a new paradigm, where the interaction *rules* that represent effective design in a stable desktop environment, may not map well to the mobile context [12]. Marshall and Tenant, for example, note four challenges for humans attempting interaction while on the go: (1) cognitive load (limited attention resources); (2) physical constraints (non-mobile activities may place constraints on physical resources); (3) terrain (external environment affects how a user will interact); and (4) other people (movement activities often involve a social element) [4].

The research to date, however, has often been acutely focused. For example, studies have been conducted to evaluate how mobile technology can be designed to compensate issues that may exist while walking and attempting text input [1, 2]. Researchers in both the practitioner and academic domains have long recognized the limitations of battery life and are constantly developing ways to extend the practical life of batteries used in smartphones [11, 13]. The inability to interact with a touch screen while wearing gloves while in a cold environment has led to the creation of specialized touchscreen gloves (i.e. [9]). All of these areas of research have in common the recognition of a specific mobile interaction issue and the addressing of that specific issue. Some recent research has begun to examine the potential need for classification of various types of situational impairments. For example, Sarsenbayeva et al. reviewed "established situational impairments" and their impact on mobile device interaction as well as suggested methods for their detection [5]. Tigwell, Flatla & Menzies examined "Situational Visual Impairments (SVI) and found that they are frequently and broadly experienced and that their root causes go beyond simply environmental sources [10].

Situational impairments have been studied as single events to be addressed through re-evaluation of how users might interact with mobile appliances. Little research to date, however, has looked into the possibility of situational impairments being comprised of complex or compound events. In addition, while some research has attempted to examine SIID classification, researchers have yet to deeply explore classification by examining a corpus of situational impairment events and creating a taxonomy from the observed events. Nor has much research attempted to explore the by-product of increased complexity that the increase in usage and functionality offered by mobile technology is engendering.

The research represented in this paper is an attempt to add these elements to the study of SIIDs. The contribution of this research is the offering of guidelines by which design of mobile human-computer interaction can (1) recognize the new complexity of the diverse facets that present during mobile interaction and (2) properly and effectively account for the presence of SIID and SCSI phenomenon in the design of mobile device interaction. The following sections provide brief outlines of the studies conducted to date as well as ongoing and planned future work for this research arc.

**STUDY 1: DIARY STUDY TO CLASSIFY SIID AND DEFINE SCSI**

In 2017, an exploratory diary study was conducted with three separate participant cohorts to better understand the types of issues faced by users of smartphone technology [6].

**Table 1: Five Main SIID Themes**

| Category | Description |
|---|---|
| **Technical Issues** | A technical fault, glitch, or other non-user or environmental issue that prevents effective completion of a transaction. |
| **Ambient Environmental Issues** | Anything about the environmental context of the transaction space that is hindering or preventing effective transaction completion. |
| **Workspace/Location Issues** | Issues that hinder or prevent the ability to effectively complete a transaction that are geospatial in nature. Either the workspace area is of insufficient size or the resources required are not within sufficient proximity to permit the effective completion of the transaction. |
| **Complexity Issues** | Issues that hinder or prevent effective transaction completion resulting from task or ambient complexity. |
| **Social/Cultural Issues** | These issues offer no physical barrier to transaction completion but nevertheless can hinder or prevent effective transaction completion. |

Each cohort participated in a session where they were introduced to the study as well as the concept of a situational impairment. The group was then informed that they were to, over a two-week period, record (when safe to do so) any time they wished interaction with their smartphone but some aspect of their current situation impacted/prevented them from completing the process to their satisfaction.

The results of the research based on the initial inquiry revealed: (1) There are at least five generalizable themes that can be used to classify situational impairment events, each with varying implications for mobile device interaction (see Table 1); and (2) A special severely constraining subset of SIID, where the multitude and complexity of ambient agents contributing to mobile I/O transaction disruption was found and defied conventional classification. These were dubbed "Severely Constraining Situational Impairments (SCSI)."

The analysis revealed some macro-characteristics of SCSIs. Due to their complexity and/or the multitude of modalities being affected during their onset, a SCSI's presence usually leads to mobile I/O transaction failure. In addition, their presence seemed to create a condition of "pre-abandonment" in some users where, based on similar past mobile I/O transaction attempts, the user perceives the current attempt as resulting in transaction failure or severe difficulty and frustration, and therefore chooses not to make an attempt, even if a successful transaction would have added some value. Still others seemed to suggest that mobile I/O transactions have a "half-life" (a period where transaction completion has value but that value is diminishing due to the passage of time). If an environmental issue is present during the need to complete a mobile I/O transaction, and that issue cannot be resolved during the period of time where the half-life has a value greater than zero, the result is a form of transaction failure as completion no longer has meaning.

While SCSIs were identified and defined, the workarounds employed by users, if any, were not fully elucidated, and design ideas for overcoming them were not explored. In addition, the SIID theme identified as "Social/Cultural" was unique relative to the other themes in that transaction impairment was exclusively the result of user volition. For example, consider an SIE that might be described as "Received a text, but cannot look at it because I am in Church". Nothing physically is preventing transaction completion, but the transaction is impaired nevertheless because the user voluntarily chose to forgo/postpone the transaction attempt. Based on the responses from the participant diaries, the researchers were able to hypothesize that there were three possible reasons for the user choosing to not complete the transaction attempt: (1) Fear of reprisal from an authority; (2) Acceptance of social/cultural norms; (3) Concern for safety. The scope of the study however, produced a limitation in being able to accurately measure motivation. In other words, the study accurately identified user volition as what caused the transaction impairment, but not why the choice was made to forego/postpone the transaction.

An additional study, therefore, was needed to better identify the motivation behind the choices that were made and how design might account for them. In addition, further research was necessary to understand whether, when a user is considering how to complete a transaction while impaired, workarounds have been developed to compensate for the lack of a technological solution. Do these workarounds suggest what the possible technological solutions should be and are they different when encountering an SIID vs. an SCSI?

**STUDY 2: INTERVIEWS, SCENARIO CREATION, PARTICIPATORY DESIGN WORKSHOPS.**

A second study was conducted consisting of two stages. A portion of the research was published as a late breaking work in 2018 [7]. During the first stage 20 smartphone users were interviewed about their habits regarding their smartphone usage, workarounds they deploy when encountering an impairment to mobile I/O transaction completion, and their underlying motivations for the choices made during the onset of a situational impairment where nothing is physically preventing transaction completion. The results of the interviews revealed a small corpus of workarounds that users attempt to deploy during the onset of a situational impairment. In addition, all 20 users revealed a compulsion to complete mobile I/O transactions even if the steps needed are, at times, putting themselves (and/or others) in potential danger. This danger is faced with, not only full acknowledgement of the physical and/or social/cultural risks, but also with a healthy acknowledgement and even acceptance of the social/cultural norms that prohibit such behavior (e.g. I know it is wrong, but there are situations where I want/need to do it anyway). While the existence of this compulsion (aka. "Nomophobia") is no longer considered novel, what was somewhat startling was that 100% of the participants willingly have at times risked danger to complete a mobile transaction.

The findings suggest that users are finding (or at least attempting to find) ways to overcome the limitations placed on them when attempting some mobile transactions in the wild. However, because being able to access data on the go has quickly become an embedded part of mobile device users' lives and common everyday activities, humans may, universally, literally be killing themselves just to read an SMS. Perhaps as is the mantra of the medical profession, to not address this reality in some way, the mobile interaction community is not adhering to the principle of "first do no harm".

As the result of analyzing the interviews, a set of three situational impairment scenarios (each with a SIID and SCSI version) was created which formed the starting point for the next stage of the study. Stage 2 consisted of a series of participatory design workshops conducted with the aim of producing actionable implications for mobile interaction design that will help users to safely and effectively perform tasks and/or gain critical information when in the presence of ambient events that are so severe and constraining that no reasonable solution attempt offers any value.

The feedback suggested through the participatory design sessions point towards an almost universal need to help address SIID/SCSI events or at least provide a way to offer support for them (e.g. providing notification reminders to do a task later on) which by inference is suggesting that this is not presently being achieved at a level that maximizes the user experience.

In addition, when asked for how they would like technology to address the common situational impairment events represented in the three scenarios, users demonstrated clear differences in the design suggestions when expressing issues associated with the onset of SCSIs. For a standard SIID, users pointed to existing technology solutions (or slight tweaks to existing technology) or even the acknowledgment of transaction postponement until the ambient conditions were more favorable (and transaction "half-life" was not an issue). There were significant differences, however, in the way that users needed technology to address the onset of a *SCSI'fied* version of the common situational impairment events. Neither existing technology, nor any non-technological workarounds were perceived as resolving the SCSI conditions. There was a common theme in all of the solutions, if technology is to best serve the user during the onset of an SCSI, technology needs to become more context aware so that it can, among other things, reduce the cognitive load required to successfully negotiate the holistic totality of the situation.

**FUTURE WORK**

The two previous studies produced important implications for design, but not specific actionable design guidelines that can be offered to designers of mobile device interaction. The first study revealed the existence of SCSIs, and that they represent a genuine accessibility challenge for mobile device use while on the go. The second study revealed that users are attempting to overcome current design limitations, sometimes at any cost, in order to complete mobile transactions in a timely manner and that the challenges that are more severely constraining might require different design considerations than that of their nominally constraining siblings.

To create a truly targeted set of design guidelines for the future, a study will be created that (1) will look for a consensus of action from the past to see if a set of principles and guidelines can be curated that will effectively map to the issues presented by the onset of SIID and SCSI events and; (2) will utilize domain experts to confirm that what has been phenomenologically gleaned from the consensus actions can be effectively mapped to address the issues that have been defined as the result of the research that has been completed to date. In addition, speaking with domain experts may also result in extending existing guidance.

**CONCLUSION**

If one defines *disruptive technology* as a technology at the core of a change in the way business is done [3], smartphones are clearly an example of this type of paradigm-altering disruption. What began to emerge as the result of conducting the research represented in the first two studies, was the realization that, like in the past, as a disruptive technology advances beyond the early adoption stage and begins to gain acceptance among the general population, new value is being added to the lives of that general population, sometimes in ways that were inconceivable pre-disruption. When adoption occurs at a rapid, exponential pace, however, negative aspects of the technology can emerge that can have detrimental effects on individual users as well as society until ways of addressing those negative aspects can be found. SCSIs are the representative of this new complexity.

The goal of this research, therefore, is to return to society a means by which we can better understand how to harness the tremendous power of mobile device interaction to enrich our lives and experiences, while reducing or even eliminating any harmful effects. This research will offer specific, actionable, curated ways for mobile design to view the new byproducts of this brave new world of mobile device interaction.